\documentclass[10pt,conference]{IEEEtran}
\IEEEoverridecommandlockouts

\usepackage{amsmath,amssymb,amsfonts}
\usepackage{graphicx}
\usepackage{textcomp}
\def\BibTeX{{\rm B\kern-.05em{\sc i\kern-.025em b}\kern-.08em
    T\kern-.1667em\lower.7ex\hbox{E}\kern-.125emX}}

\usepackage{cite}
\usepackage{enumitem}
\usepackage{booktabs}
\usepackage[table]{xcolor}  
\usepackage{colortbl}
\usepackage{pifont}
\usepackage[most]{tcolorbox}
\usepackage[numbers]{natbib}
\usepackage[normalem]{ulem}
\usepackage{multirow}
\usepackage{subcaption}   
\usepackage{algorithm}
\usepackage{algpseudocode}
\usepackage{xurl}
\usepackage[framemethod=tikz]{mdframed}

\definecolor{lightgrey}{rgb}{0.93,0.93,0.93}

\usepackage[framemethod=tikz]{mdframed}
\mdfdefinestyle{customstyle}{
  linecolor=gray!100,
  linewidth=3pt,
  innerleftmargin=3pt,
  topline=false,
  rightline=false,
  bottomline=false,
  leftline=true,
  innerrightmargin=3pt,
  innertopmargin=3pt,
  innerbottommargin=3pt,
  backgroundcolor=gray!15
}

\newenvironment{custommdframed}
  {\begin{mdframed}[style=customstyle]}
  {\end{mdframed}}

\newcommand{\ourmethod}{\textsc{EvoCodeAgent}}

\begin{document}

\title{\ourmethod{}: Project-Level Code Generation via Architecture-Code Co-Evolution and Experience-Guided Repair}

\author{{\bf Qianhui Zhao}, {\bf Li Zhang}, {\bf Fang Liu}$^{*}$ \\ 
School of Computer Science and Engineering, \\ State Key Laboratory of Complex \& Critical Software Environment,\\ Beihang University\\
\texttt{\{zhaoqianhui, fangliu\}@buaa.edu.cn}
}



\maketitle

\begin{abstract}
Project-level code generation aims to automatically synthesize complete software repositories from high-level requirements, producing coherent multi-file systems with well-structured modules and consistent cross-file dependencies. Despite the strong capabilities of large language models (LLMs) in specification understanding and code synthesis, existing approaches still exhibit several critical limitations.
First, many methods rely on static architectural designs that cannot adapt during iterative generation, leading to a growing misalignment between architecture and implementation. Second, they lack mechanisms to accumulate and reuse experience across iterations, resulting in recurring errors and reduced efficiency. Third, they often depend on standalone LLM-based evaluation of intermediate artifacts, which suffer from limited transparency and potential unreliability under long-context reasoning.
To address these challenges, we propose \ourmethod{}, a state-centric framework for project-level code generation.
\ourmethod{} maintains an evolving architectural state initialized from structured project specifications and continuously reconciles predicted relations from architecture and skeletons with realized relations extracted from generated code.
It further uses tool-grounded verification evidence to support reliable state updates and abstracts reusable repair experiences from previous revision attempts to guide subsequent generation.
Extensive experiments on project-level benchmarks of different scales, including DevBench and CodeProjectEval, demonstrate that \ourmethod{} consistently outperforms baseline approaches. Under the GPT-5 setting, it achieves unit test pass rates of 60.48\% and 27.73\%, respectively, yielding up to a 53.5\% improvement over the strongest baseline.
\end{abstract}


\section{Introduction}

Project-level code generation aims to automatically synthesize an entire project from high-level requirement descriptions, including its file organization and executable implementations \cite{li2025prompting, zan2024codes, zan2023large}. In real-world software engineering, developers typically build complete software repositories consisting of multiple files, modules, and interdependent components rather than isolated code snippets. As software development increasingly relies on automated assistance to improve productivity and reduce engineering effort \cite{wei2022chain, yao2022react, gao2025trae, wang2024openhands, austin2021program, shinn2023reflexion}, enabling large language models (LLMs) to generate project-scale systems has become an important research goal. Compared with traditional code generation tasks that focus on producing individual functions or small code fragments based on localized contexts \cite{chen2021codex, austin2021mbpp, liu2024repobench, zhang2023repocoder, yu2024codereval}, project-level code generation requires LLMs to construct coherent multi-file systems with well-organized modules and consistent cross-file dependencies. Consequently, it represents a substantially more challenging yet practically important problem in automated software engineering.

Recent research has made progress in project-level code generation. 
As state-of-the-art LLMs such as Claude \cite{anthropic2025claude4}, DeepSeek \cite{guo2024deepseek, liu2024deepseekv3, guo2025deepseekr1}, Gemini \cite{team2023gemini, team2024gemini}, and GPT \cite{hurst2024gpt, achiam2023gpt} demonstrate strong capabilities in understanding natural language specifications and generating high-quality code, LLM-based approaches \cite{hong2023metagpt,qian2024chatdev,zan2024codes, zhao2025projectgen, seo2025paper2code, luo2025rpg, liu2025evodev, nguyen2025agilecoder} have become the main force in this task.
Despite these encouraging results, existing approaches still face several important limitations:

\ding{182} \textbf{Static architectural representations cannot capture architectural evolution throughout the implementation process.}
Many existing approaches \cite{zan2024codes, hong2023metagpt, qian2024chatdev, seo2025paper2code} directly generate code from requirement descriptions without explicitly introducing an architectural design stage, which may suffer from the semantic gap between natural language requirements and executable code.
To mitigate this issue, some recent methods \cite{zhao2025projectgen, luo2025rpg} incorporate structured intermediate representations as architectural guidance.
However, these representations are typically constructed at an early planning stage and are mainly used as fixed blueprints during subsequent generation.
Such a design is inherently limited, as architectural specifications produced before implementation may be incomplete or inaccurate due to ambiguous requirements and missing low-level details.
As generation proceeds, skeletons may introduce new imports, implementations may revise interfaces, and execution feedback may reveal missing or inconsistent components.
These implementation-derived facts naturally require the architectural representation to be revisited and updated.
Without continuously reconciling predicted architectural relations with realized implementation relations, later generation steps may rely on outdated dependencies or incomplete interfaces, leading to architecture-code drift.

\ding{183} \textbf{Iterative feedback lacks reusable experience abstraction.}
Project-level code generation typically involves multiple rounds of iterative refinement, during which LLMs may repeatedly encounter similar issues, such as missing dependencies, inconsistent interfaces, import failures, or incomplete implementations.
However, existing approaches \cite{zhao2025projectgen} mainly use feedback, local diffs, or recent revision histories as contextual memory for the current iteration.
Although such information can support local fixes, it is often tightly coupled with specific files, error traces, and code contexts.
As a result, when similar issues reappear in later stages or different files, LLMs may struggle to transfer prior repair knowledge and may repeatedly make the same mistakes.
Without mechanisms to abstract concrete refinement trajectories into context-independent repair scenarios and strategies, iterative feedback may not effectively guide subsequent generations beyond the immediate local fix.

\ding{184} \textbf{Standalone LLM-based judgments provide limited reliability for intermediate artifact validation.}
Multi-stage code generation frameworks typically produce a series of intermediate artifacts that should be validated before proceeding to subsequent steps.
In many existing approaches \cite{hong2023metagpt, qian2024chatdev, nguyen2025agilecoder, zhao2025projectgen}, this validation is predominantly handled by LLM-based judges in a standalone manner, without sufficient support from external tools or structured verification mechanisms.
In practice, these evaluations often require LLMs to assess large-scale natural language descriptions or code artifacts under complex long-context settings.
Due to inherent limitations in long-context reasoning, such as context dilution and attention degradation, LLMs may fail to consistently incorporate all relevant information \cite{liu2026beyond, zhang2025llmhallucinations}, leading to unstable or inconsistent judgments.
Moreover, standalone judgments usually provide limited verifiable evidence about which files, interfaces, dependencies, or requirements are missing or inconsistent.
This lack of actionable evidence makes the results difficult to verify and limits their usefulness for state updates and targeted repair.

To address these limitations, we propose \ourmethod{}, a state-centric project-level code generation framework.
Instead of treating artifact as a static intermediate representation, \ourmethod{} maintains an evolving project state throughout generation.
The state is initialized from project requirements and continuously updated throughout generation using verification feedback, allowing architecture to guide code generation while enabling generated artifacts to refine architectural decisions.
In addition, \ourmethod{} grounds intermediate evaluation with specialized verification tools, producing explicit evidence for state reconciliation and repair.
Finally, it abstracts reusable repair experiences from refinement trajectories, allowing recurring error patterns to be addressed with accumulated knowledge rather than rediscovered from scratch.
Experimental results demonstrate that \ourmethod{} achieves state-of-the-art performance on project-level code generation tasks. Under the GPT-5 setting, \ourmethod{} attains a unit test pass rate of 60.48\% on the small-scale DevBench benchmark and 27.73\% on the larger-scale CodeProjectEval benchmark, yielding up to a 53.5\% relative improvement over the strongest baseline.

Our contributions can be summarized as follows:

\begin{itemize}
    \item We identify key limitations in existing project-level code generation frameworks, including static architecture-code misalignment, limited reuse of iterative repair experience, and unreliable standalone LLM-based evaluation.

    \item We propose \ourmethod{}, a state-centric project-level code generation framework.
    It maintains an evolving project state to support architecture-code co-evolution, tool-grounded verification, and experience-guided repair.

    \item We conduct extensive experiments on project-level benchmarks of different scales, including DevBench and CodeProjectEval. 
    Results show that \ourmethod{} consistently and significantly outperforms strong baselines under various experimental settings. Our code can be found at \url{https://github.com/whisperzqh/EvoCodeAgent.git}.
\end{itemize}

\section{Related Work}

\subsection{Project-level Code Generation Benchmarks}

Project-level code generation aims to synthesize a complete multi-file software project from high-level requirements \cite{li2025prompting}, requiring the construction of interdependent components while maintaining architectural consistency and cross-file dependencies. This differs from repository-level function generation, where models generate individual functions given repository context \cite{zhang2023repocoder, wang2024rlcoder, zhang2024codeagent, wu2024repoformer, yu2024codereval}.

Early datasets such as SRDD \cite{qian2024chatdev}, SRLCG \cite{ma2025srlcg}, CASSD \cite{zhang2024experimenting}, and SoftwareDev \cite{hong2023metagpt}, provide only software requirements without ground-truth code, leading to evaluations based on heuristic metrics or human judgment, rather than direct comparison with reference implementations.
Later benchmarks introduce ground-truth repositories to enable more objective evaluation. SketchEval \cite{zan2024codes} measures project-level similarity via SketchBLEU, but such similarity-based evaluation cannot fully capture functional correctness. DevBench \cite{li2025prompting} incorporates executable test cases across multiple languages, though its projects remain relatively small (2.2 files and 276 LOC on average).
Commit0 \cite{zhaocommit0} and JavaBench \cite{cao2024javabench} further provide executable evaluation settings with starter code or code skeletons.
Commit0 covers 54 Python libraries, while JavaBench extends project-level evaluation to Java and object-oriented programming with four Java projects, 106 classes, and 389 methods.
However, both focus on completing missing implementations within provided code structures, rather than generating complete repositories from scratch.
More recent benchmarks emphasize larger scale and realism. CodeProjectEval \cite{zhao2025projectgen} offers 18 fully runnable repositories with richer structures (12.7 files and 2,388 LOC on average) and comprehensive tests, enabling reliable evaluation. 
NL2Repo-Bench \cite{ding2025nl2repo} further extends to long-horizon generation with 104 tasks across diverse domains, varying difficulty levels.

\begin{figure*}[t]
    \centering
    \setlength{\abovecaptionskip}{0.1cm}
    \includegraphics[width=\linewidth]{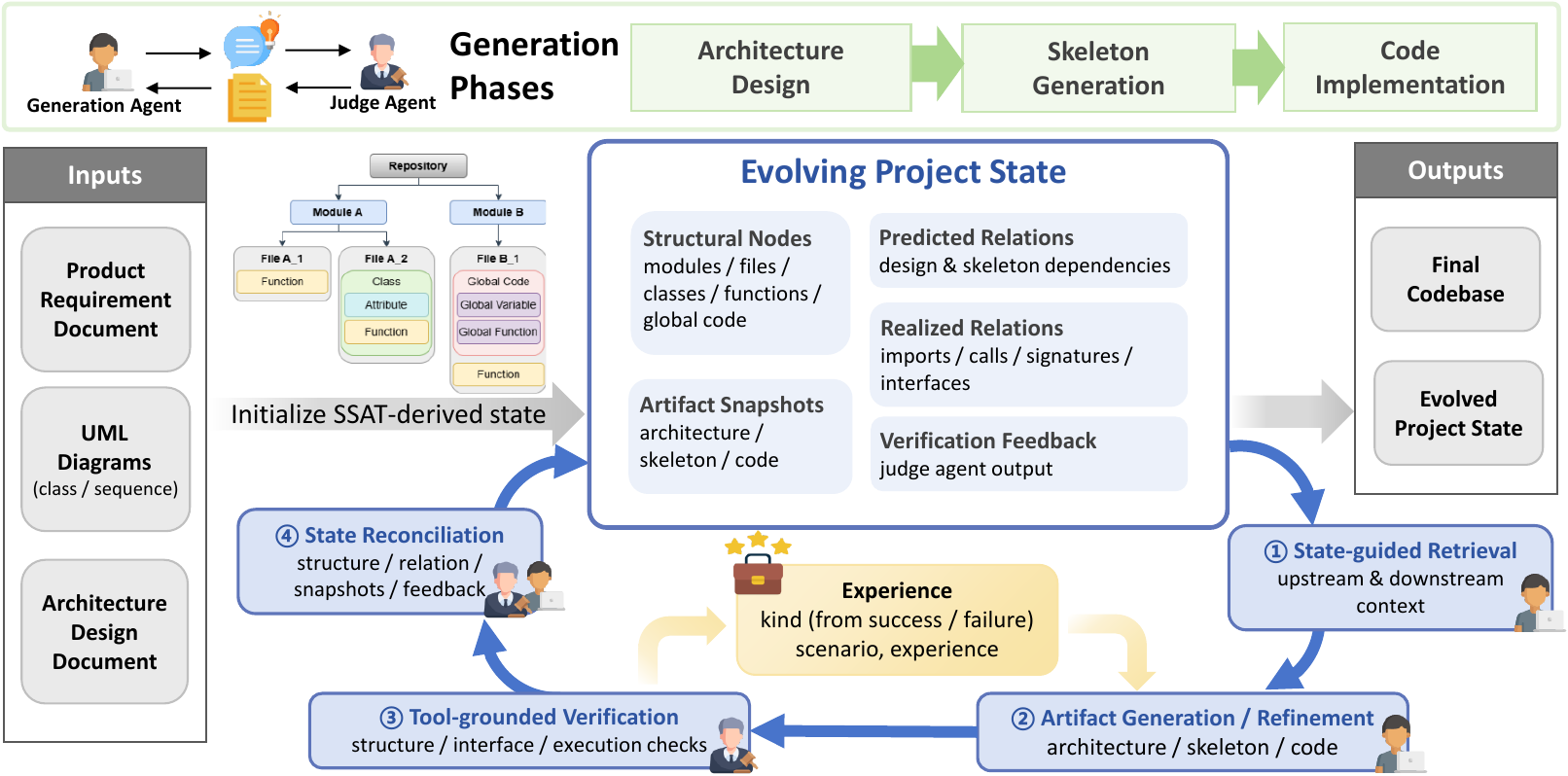}
    \caption{Overview of \ourmethod{}.}
    \label{fig:approach}
    \vspace{-0.3cm}
\end{figure*}

\subsection{LLM-based Project-level Code Generation Approaches}

Existing approaches to project-level code generation can be broadly categorized into pipeline-based methods and agent-based frameworks.

Pipeline-based approaches break the generation into sequential stages. 
CodeS \cite{zan2024codes} adopts a hierarchical strategy that progressively generates directory structures, file-level sketches, and function implementations. 
SRLCG \cite{ma2025srlcg} follows a similar paradigm, utilizing a multidimensional chain-of-thought and self-rectification, and integrating generated files through dynamic backtracking. 
While these methods produce structured outputs, they largely rely on one-pass generation with limited intermediate validation, making them prone to error propagation across stages.

Agent-based frameworks coordinate multiple LLM-driven agents to simulate software development workflows.
MetaGPT \cite{hong2023metagpt} organizes role-based agents within a global memory pool, whereas ChatDev \cite{qian2024chatdev} enables multi-turn collaboration among specialized agents through dialogue-based collaboration.
However, both primarily target small-scale tasks and suffer from limited long-context memory.
AgileCoder \cite{nguyen2025agilecoder} introduces role-based iterative development with a dynamically maintained dependency graph for context retrieval and testing, but this graph depends on already generated code and thus provides limited guidance in early stages.
ZeroRepo \cite{luo2025rpg} constructs a Repository Planning Graph (RPG) for graph-guided repository generation.
Although RPG is progressively constructed during planning, it does not explicitly reconcile planned relations with implementation relations extracted from generated code during generation.
ProjectGen \cite{zhao2025projectgen} further structures generation into architecture, skeleton, and code stages, introducing the Semantic Software Architecture Tree (SSAT) to model modules and dependencies.
Similarly, the SSAT remains static after construction, and its reliance on solely LLM-based judgment may lead to unreliable evaluation.

In contrast, \ourmethod{} treats the architectural representation as an evolving project state rather than a static planning artifact.
It continuously reconciles predicted relations inferred from architecture and skeletons with realized relations extracted from generated code, while using tool-grounded evidence and reusable repair experiences to guide state updates and subsequent generation.

\section{Approach}
\label{sec:approach}

To support adaptive and reliable project-level code generation, we propose \ourmethod{}, a state-centric framework that combines staged multi-agent generation with continuous state evolution.
As shown in Figure~\ref{fig:approach}, \ourmethod{} proceeds through three phases: architecture design, skeleton generation, and code implementation.
Each phase is managed by a generation--judge agent pair, where the generation agent produces or revises phase-specific artifacts and the judge agent evaluates them with feedback, enabling iterative refinement within the phase.

Rather than following a one-way staged pipeline, \ourmethod{} connects the three phases through an evolving project state initialized from the input artifacts, including the product requirement document, UML diagrams, and architecture design document.
During generation, this state is updated through state-guided retrieval, artifact generation, tool-grounded verification, and state reconciliation.
Thus, architecture guides downstream skeleton and code generation, while generated artifacts and verification feedback are written back to refine earlier architectural assumptions.
In addition, execution feedback from the code implementation stage is distilled into reusable experience memory, providing guidance for subsequent iterations.
The following subsections describe these mechanisms in detail.

\subsection{Evolving Project State}
\label{sec:evolving-state}

The evolving project state serves as the core representation in \ourmethod{}. It extends the SSAT \cite{zhao2025projectgen}, which provides a machine-parsable hierarchy of modules, files, classes, functions, and global code blocks, into a richer dynamic representation. Beyond the static structural hierarchy encoded by SSAT, the project state additionally maintains dependency relations, generated artifacts, and verification feedback, and is continuously updated throughout generation. The reusable experience set is maintained alongside the project state and is described in Section~\ref{sec:experience}.

Formally, the project state at iteration $t$ is defined as:
\begin{equation}
P_t = (N_t, R_t^{p}, R_t^{r}, A_t, V_t),
\label{equation: state}
\end{equation}
where $N_t$ denotes structural nodes, $R_t^{p}$ denotes predicted relations, $R_t^{r}$ denotes realized relations, $A_t$ denotes generated artifact snapshots, and $V_t$ denotes verification feedback.

Specifically, $N_t$ records the SSAT-derived hierarchical structure of the project, including modules, files, classes, functions, global code blocks, and their containment relations. $R_t^{p}$ and $R_t^{r}$ denote predicted and realized dependencies, respectively, providing architectural priors before implementation and updated dependency information during code implementation. By continuously reconciling predicted and realized relations, \ourmethod{} keeps architectural guidance aligned with the evolving codebase. $A_t$ stores snapshots of generated architecture, skeleton, and code artifacts, enabling comparison across iterations. $V_t$ stores structured verification feedback produced by judge agents based on tool checks and test execution, such as missing files, uncovered interfaces, failed check tests, and repair targets.

Together, these components provide a unified foundation for dependency-aware retrieval, state reconciliation, and iterative refinement.

\subsection{Architecture-Code Co-Evolution}
\label{sec:coevolution}

\begin{algorithm}[t]
\footnotesize
\caption{Architecture-Code Co-evolution}
\label{alg:coevolution}
\begin{algorithmic}[1]

\State \textbf{Input:} Initial architecture state $\mathcal{P}_a$, max skeleton iterations $T_s$, max code iterations $T_c$
\State \textbf{Output:} Final codebase $\mathcal{C}^{(T_c)}$ and evolved project state $\mathcal{P}_c^{(T_c)}$

\State \textbf{Skeleton Generation Phase}
\State $\mathcal{P}_s^{(0)} \gets \mathcal{P}_a$

\For{$t_s = 1$ to $T_s$}
    \State Retrieve repair targets from verification feedback
    \State Generate or refine partial skeleton $\Delta \mathcal{S}^{(t_s)}$

    \State Update project state: 
    \[
        \mathcal{P}_s^{(t_s)}
        \gets
        \mathrm{ReconcileState}
        (\mathcal{P}_s^{(t_s-1)},
        \Delta \mathcal{S}^{(t_s)},
        \Delta \mathcal{V}^{(t_s)})
    \]
\EndFor

\State Infer predicted relations:
$
    \mathcal{R}^{p}
    \gets
    f_{\mathrm{infer}}(\mathcal{P}_s^{(T_s)})
$

\State Initialize code generation state:
\[
    \mathcal{P}_c^{(0)}
    \gets
    \mathrm{UpdateRelations}
    (\mathcal{P}_s^{(T_s)}, \mathcal{R}^{p})
\]

\State \textbf{Code Implementation Phase}
\For{$t_c = 1$ to $T_c$}
    \State $\mathcal{P}_c^{(t_c)} \gets \mathcal{P}_c^{(t_c-1)}$

    \For{each file node $n_i \in \mathcal{N}^{\mathrm{file},(t_c)}_c$}

        \State Retrieve dependency-aware context:
        \[
            \mathcal{C}_{\mathrm{ctx}}(n_i)
            \gets
            \mathrm{RetrieveContext}
            (n_i, \mathcal{P}_c^{(t_c)})
        \]

        \State Generate or refine code:
        \[
            \mathcal{C}^{(t_c)}(n_i)
            \gets
            f_{\mathrm{gen}}(\mathcal{C}_{\mathrm{ctx}}(n_i))
        \]

        \State Extract realized structures:
        \[
            \Delta \mathcal{N},
            \Delta \mathcal{R}^{r}
            \gets
            f_{\mathrm{extract}}
            (\mathcal{C}^{(t_c)}(n_i))
        \]

        \State Reconcile project state:
        \[
            \mathcal{P}_c^{(t_c)}
            \gets
            \mathrm{ReconcileState}
            (\mathcal{P}_c^{(t_c)},
            \Delta \mathcal{N},
            \Delta \mathcal{R}^{r},
            \Delta \mathcal{V})
        \]

    \EndFor
\EndFor

\State \Return $\mathcal{C}^{(T_c)}, \mathcal{P}_c^{(T_c)}$

\end{algorithmic}
\end{algorithm}

\begin{table*}[t]
\centering
\setlength{\abovecaptionskip}{0.1cm}
\caption{Task-oriented tools in \ourmethod{}.}
\label{tab:tools}
\resizebox{\linewidth}{!}{
    \begin{tabular}{lll}
    \toprule
    \textbf{Tool Name} & \textbf{Description} & \textbf{Technical Basis} \\ 
    \midrule
    \rowcolor{lightgrey}
    \multicolumn{3}{l}{\textbf{Tools for Generation Agents}}\\
    Requirement Access & Access requirement artifacts from PRD, UML, and architecture documents & Hierarchical document access \\ 
    Symbol Extractor & Extract all classes, functions, and variable symbols from project state nodes & Structure-aware traversal, symbol extraction \\ 
    Artifacts Retriever & Retrieve project artifacts (architecture nodes, skeletons, or implemented code) for a given file path & Path-based indexing \\ 
    \midrule
    
    \rowcolor{lightgrey}
    \multicolumn{3}{l}{\textbf{Tools for Judge Agents}}\\
    Requirement Access & Access requirement artifacts from PRD, UML, and architecture documents & Hierarchical document access \\ 
    Artifacts Retriever & Retrieve project artifacts (architecture nodes, skeletons, or implemented code) for a given file path  & Path-based indexing\\ 
    Documentation Searcher & Retrieve the most relevant blocks from project documentation based on keyword queries & Keyword search, semantic relevance ranking \\
    Structure Verifier & Verify artifacts against structure node specifications for structural and property completeness & Rule-based hierarchical validation \\ 
    Interface Verifier & Verify architecture completeness against UML-defined classes, functions, attributes, and global blocks & UML parsing, node extraction and matching\\ 
    Missing File Detector & Detect code files defined in the architecture but absent in architecture nodes & nodes-guided file existence verification \\ 
    Skeleton Coverage Verifier & Verify skeleton code includes all architecture-specified files, classes, and functions & AST-based code extraction, architecture node matching \\ 
    \bottomrule
    \end{tabular}
    }
    \vspace{-0.3cm}
\end{table*}

\ourmethod{} first initializes the architecture state $\mathcal{P}_a$ by analyzing the input artifacts. This produces the structural nodes $\mathcal{N}_a$, architecture artifacts $\mathcal{A}_a$, and architecture-level verification feedback $\mathcal{V}_a$. At the end of the architecture design phase, both predicted and realized relations remain empty, since only structural containment information is available before skeleton generation.

To bridge the gap between architectural design and implementation in conventional staged generation, \ourmethod{} performs architecture-code co-evolution over the evolving project state.
Specifically, dependency relations are progressively constructed and refined during skeleton generation and code implementation, allowing architectural guidance to remain aligned with the evolving codebase.
Algorithm~\ref{alg:coevolution} summarizes this process.

During the skeleton generation phase, the agent iteratively generates and refines skeleton artifacts based on the current skeleton state.
After each refinement step, judge agents perform structural verification to identify missing interfaces, inconsistent declarations, and other structural issues.
The refined skeleton together with the newly produced verification feedback is reconciled into the project state, enabling progressive correction of structural inconsistencies.
After the final skeleton iteration, the system analyzes the complete skeleton to infer a set of \emph{predicted relations}, derived from multiple signals such as import statements, cross-file symbol matching, and semantic similarity between code elements.
These predicted relations capture potential dependencies not explicitly encoded in the initial structural nodes and provide dependency priors for subsequent code generation.

During code implementation, the evolving project state guides context retrieval for each target file, enabling dependency-aware and structure-consistent generation.
For a target file node $n_i$, incoming and outgoing dependencies are identified based on the current relation state.
In the first code iteration, context retrieval primarily relies on predicted relations $\mathcal{R}^{p}$, since implementation-derived dependencies are not yet available.
In later iterations, realized relations $\mathcal{R}^{r}$ extracted from generated code progressively dominate dependency retrieval, while predicted relations serve as supplementary priors.
The retrieved artifacts are aggregated into context $\mathcal{C}_{\mathrm{ctx}}(n_i)$ and provided to the generation agent for implementation.
This dependency-aware retrieval ensures that both upstream dependencies (e.g., imported modules) and downstream usages (e.g., calling sites) are considered, improving interface consistency and cross-file interaction correctness.

After code generation, the implementation is evaluated through test execution to obtain structured feedback, such as interface mismatches and failed check tests.
Static analysis further extracts \emph{realized relations} and structural updates that reflect actual dependencies in the generated code.
These signals are reconciled into the project state, continuously aligning architectural assumptions with implementation reality.
Through iterative state reconciliation, the co-evolution mechanism maintains a bidirectional linkage between architecture and code, providing a unified and adaptive backbone for project-level code generation.

\subsection{Tool-Grounded Verification}
\label{sec:tool-verification}

\ourmethod{} equips both generation and judge agents with task-oriented tools to enable selective context retrieval and structured artifact verification, as summarized in Table~\ref{tab:tools}. Rather than relying solely on LLM reasoning, \ourmethod{} adopts a hybrid verification paradigm in which judge agents combine structured tool outputs with LLM-based reasoning to produce more reliable and interpretable feedback.

Generation agents primarily use tools to retrieve compact and relevant context, including requirement artifacts, symbols from the current project state, and previously generated artifacts, thereby reducing irrelevant context and improving reasoning efficiency. Judge agents use tools to collect explicit verification evidence, such as missing architectural elements, uncovered interfaces, and structural inconsistencies, which serves as the basis for iterative refinement.

The role of tools varies across different artifact granularities. At the architecture level, verification focuses on structural completeness and consistency with input artifacts, enabling the judge agent to identify missing nodes, incomplete interfaces, and architectural inconsistencies. At the skeleton level, verification ensures that the generated skeleton faithfully covers all files, classes, and functions specified in the current project state while maintaining syntactic correctness. At the implementation level, verification is primarily driven by executable feedback from predefined check tests, which provides direct evidence of functional correctness.

The structured evidence collected from all stages is written into the verification component $V_t$ of the evolving project state. These verification signals are subsequently used during state reconciliation and future generation, making tool-based verification not only a mechanism for evaluation but also a source of actionable guidance for iterative refinement. By grounding agent decisions in explicit evidence rather than relying solely on implicit LLM reasoning, this design improves verification reliability, reduces error propagation across stages, and enables more robust project-level code generation under long-context settings.

\subsection{Experience Abstraction and Reuse}
\label{sec:experience}

\ourmethod{} accumulates reusable repair experiences from iterative refinement trajectories to support long-horizon self-improvement.
Project-level code generation frequently encounters recurring issues, such as incompatible imports, inconsistent public APIs, incorrect serialization behavior, or incomplete exception handling.
Although these issues may arise in different files or at different iterations, they often share similar failure patterns and repair strategies.
If the system only retains raw feedback from previous iterations, similar errors may still need to be rediscovered and repaired repeatedly.
To address this, \ourmethod{} abstracts concrete modification trajectories into reusable repair experiences.

After each refinement iteration, \ourmethod{} compares artifact snapshots and verification outcomes before and after modification.
Let $A_{t-1}$ and $A_t$ denote artifact snapshots before and after a refinement step, and let $V_{t-1}$ and $V_t$ denote the corresponding verification evidence.
These differences capture what was modified, which issue the modification attempted to resolve, and whether the resulting verification outcome improved.

Based on these trajectory differences, \ourmethod{} extracts an experience:
\begin{equation}
e_i = (s_i, a_i, y_i),
\end{equation}
where $s_i$ denotes an abstract scenario, $a_i$ denotes the corresponding corrective strategy, and $y_i$ denotes the observed outcome.
The scenario summarizes the conditions under which the experience is applicable, such as an import-time compatibility failure or an interface mismatch between method signatures and calling sites.
The strategy describes the corresponding repair principle, such as preserving optional parameters.
The outcome records whether the strategy improved verification results, produced no observable effect, or introduced regressions.
The experience set is updated incrementally:
\begin{equation}
E_t = E_{t-1} \cup \mathrm{Abstract}(A_{t-1}, A_t, V_{t-1}, V_t).
\end{equation}

During subsequent generation or repair in code implementation stage, \ourmethod{} retrieves experiences relevant to the current context.
Experiences retrieved by the error messages are injected into the generation context as high-level guidance rather than direct code patches, allowing the generation agent to reuse repair principles while adapting them to the current implementation.

This mechanism differs fundamentally from storing local historical feedback.
Raw feedback and code diffs are often tied to specific files or iterations, whereas experiences in \ourmethod{} are abstracted into reusable knowledge associated with explicit outcomes.
As a result, they can generalize across files and later iterations when similar scenarios arise.
By transforming past refinement trajectories into reusable guidance, \ourmethod{} reduces repeated error discovery, improves repair efficiency, and enables more effective iterative self-correction during project-level code generation.

\section{Evaluation}

To evaluate the effectiveness of \ourmethod{}, we address the following research questions:

\vspace{-0.15cm}
\begin{center}\small
\begin{tcolorbox}[colback=gray!10,
                  colframe=black,
                  arc=1mm, auto outer arc,
                  boxrule=0.5pt, breakable
                 ]
\textbf{RQ1: Overall Performance.}
How does the generation performance of \ourmethod{} compare against state-of-the-art approaches on project-level code generation benchmarks?

\textbf{RQ2: Ablation Study.}
What are the contributions of each component of \ourmethod{} to the overall performance? 

\textbf{RQ3: Efficiency Analysis.}
How does \ourmethod{} balance quality, cost, and token efficiency against existing approaches?

\end{tcolorbox}
\end{center}

\subsection{Datasets and Backbone Models}

To support automated test-based evaluation, we use two datasets, DevBench \cite{li2025prompting} and CodeProjectEval \cite{zhao2025projectgen}, both providing PRDs, UML diagrams, and architectural specifications for each task.

\begin{itemize}[leftmargin=20pt]
    \item \textbf{DevBench}: It is designed to evaluate LLM performance across the full software development lifecycle. We use the Python subset of DevBench, which contains 10 tasks with an average of 2.2 files and 276 lines of code each.
    \item \textbf{CodeProjectEval}: It is designed to support realistic evaluation through executable test cases and structured project artifacts, containing 18 Python tasks across diverse domains with an average of 12.7 files and 2,388.6 LOC.
\end{itemize}

We conduct experiments using two backbone LLMs, DeepSeek-v3.2 \cite{liu2025deepseek} and GPT-5 \cite{singh2025gpt5}, both of which have demonstrated strong capabilities in code generation.

\subsection{Baselines}

We compare \ourmethod{} with three state-of-the-art project-level code generation approaches. All of these approaches take user requirements as input and directly generate complete project-level code as output. Detailed information can be seen as follows:
\begin{itemize}[leftmargin=20pt]
    \item \textbf{CodeS} \cite{zan2024codes}: It introduces a hierarchical sketching framework that decomposes project-level code generation into multi-granularity sub-tasks through a structured sketching and filling process.
    \item \textbf{MetaGPT} \cite{hong2023metagpt}: It models software development as a multi-agent collaboration workflow guided by predefined standard operating procedures, enabling controllable task decomposition and information exchange among specialized agents.
    \item \textbf{ProjectGen} \cite{zhao2025projectgen}: It is a multi-agent framework for project-level code generation that decomposes the process into architecture design, skeleton generation, and code completion, using a semantic software architecture tree and iterative refinement with memory-based context management to bridge requirements and implementation.
\end{itemize}

\begin{table*}[t]
    \centering
    \small
    \setlength{\abovecaptionskip}{0.1cm}
    \caption{Comparison between \ourmethod{} and baseline approaches on DevBench and CodeProjectEval dataset. \textit{\%Pass} represents the overall pass rate on the dataset, while the other metrics are average / median values across individual tasks.}
    \resizebox{0.85\linewidth}{!}{
    \begin{tabular}{ll|cc|cc|cc|cc|c}
        \toprule
        \multirow{2}{*}{\textbf{Model}} & \multirow{2}{*}{\textbf{Approach}} & \multicolumn{2}{c|}{\textbf{\#Files}} & \multicolumn{2}{c|}{\textbf{LOC}} & \multicolumn{2}{c|}{\textbf{Complexity}} & \multicolumn{2}{c|}{\textbf{Arch Consistency}} & \multirow{2}{*}{\textbf{\%Pass}} \\
         & & \textbf{Avg.} & \textbf{Mid.} & \textbf{Avg.} & \textbf{Mid.} & \textbf{Avg.} & \textbf{Mid.} & \textbf{Avg.} & \textbf{Mid.}  & \\
        \midrule
        \rowcolor{lightgrey}
        \multicolumn{11}{c}{\textbf{DevBench}}\\
        \midrule
         \multirow{4}{*}{\textbf{DeepSeek-v3.2}} &
        CodeS & 3.1 & 2.5 & 563.3 & 372.0 &	5.56 & 5.82 & / & / & 25.00\% \\
         & MetaGPT & 1.7 & 1.0 & 243.0 & 260.0 & 2.48 & 2.90 & / & / & 29.84\%\\
         & ProjectGen & 3.0 & 2.0 & 303.2 & 256.5 & 3.40 & 3.20 & 58.80 & 57.91 & 42.74\%\\
         & \ourmethod{} & 3.0 & 2.0 & 331.0 & 361.5 & 3.91 & 3.63 & \textbf{78.92} & \textbf{100.0 }& \textbf{50.81\%} \\
        \midrule
        \multirow{4}{*}{\textbf{GPT-5}} &
        CodeS & 2.4 & 2.0 & 277.9 & 255.0 & 4.51 & 4.15 & / & / & 4.03\% \\
         & MetaGPT & 1.8 & 1.5 & 226.3 & 219.0 & 2.95 & 3.50 & / & / & 32.26\%\\
         & ProjectGen & 3.0 & 2.0 & 460.3 & 459.0 & 8.17 & 7.16 & 61.94 & 61.82 & 57.26\%\\
         & \ourmethod{} & 3.0 & 2.0 & 454.2 & 431.5 & 8.28 & 6.52 & \textbf{84.92} & \textbf{98.65}  & \textbf{60.48\%}\\
        \midrule
        \rowcolor{lightgrey}
        \multicolumn{11}{c}{\textbf{CodeProjectEval}}\\
        \midrule
        \multirow{4}{*}{\textbf{DeepSeek-v3.2}} &
        CodeS & 1.9 & 0.0 & 201.3 & 0.0 & 0.93 & 0.00 & / & / & 0.00\%\\
         & MetaGPT & 10.5 & 9.5 & 1838.0 & 988.0 & 4.30 & 2.20 & / & / & 1.04\% \\
         & ProjectGen & 5.5 & 5.5 & 643.1 & 619.0 & 4.32 & 2.55 & 32.70 & 31.65 & 3.16\% \\
         & \ourmethod{} & 9.3 & 8.5 & 1367.4 & 1280.0 & 3.30 & 2.80 & \textbf{46.69} & \textbf{43.19} & \textbf{7.02\%}\\
        \midrule
         \multirow{4}{*}{\textbf{GPT-5}} &
         CodeS & 6.9 & 5.5 & 1249.7 & 966.5 & 3.52 & 3.73 & / & / & 1.51\%\\
         & MetaGPT & 9.0 & 7.5 & 1559.9 & 1439.0 & 2.30 & 2.52 & / & / & 4.00\% \\
         & ProjectGen & 12.7 & 11.5 & 2468.2 & 2274.5 & 5.55 & 4.60 & 55.78 & 56.03  & 18.07\%\\
         & \ourmethod{} & 13.3 & 11.5 & 2352.4 & 2399.0 & 5.70 & 5.19 & \textbf{86.04} & \textbf{92.45}  & \textbf{27.73\%}\\
        \bottomrule
    \end{tabular}
    }
    \label{tab:main_results}
    \vspace{-0.3cm}
\end{table*}

\subsection{Metrics}

We evaluate the generated repositories using the following metrics, covering functional correctness, architectural alignment, and code complexity.

\begin{itemize}[leftmargin=20pt]

\item \textbf{Pass rate (\%Pass)}:
We use unit test pass rate as the \textit{primary metric} to evaluate code generation performance, as it directly reflects the functional correctness of the generated repository. It is computed as the percentage of unit tests successfully passed by the generated code.

\item \textbf{Architecture consistency}:
We define architecture consistency as a metric for measuring structural alignment between the generated and ground-truth repository. Specifically, we extract all root-to-leaf paths from the structural nodes $N_{a}$ at the end of architecture design phase and reconstruct the corresponding paths in the ground-truth repository via AST parsing. The metric is defined as:
\begin{equation}
ArchConsistency=\frac{2|P_n \cap P_g|}{|P_n|+|P_g|}
\end{equation}
where $P_n$ and $P_g$ denote the path sets of the generated and ground-truth repositories, respectively. Since each path captures the hierarchical dependency chain of a software component, the F1 score over path sets provides a balanced measure.

\item \textbf{Number of generated files (\#Files)}:
We record the number of generated files to assess the model's ability to accurately reconstruct the project structure.

\item \textbf{Lines of code (LOC)}:
We compute the total number of lines of generated code to estimate the scale of produced implementations and analyze differences in generation completeness across methods.

\item \textbf{Code complexity}:
We measure the cyclomatic complexity of the generated code to assess its structural quality and maintainability.

\end{itemize}

\subsection{Implementation Details}

For \ourmethod{}, we implement the multi-agent system using LangChain\footnote{https://www.langchain.com}, a framework for building applications with large language models.
To retrieve relevant experience, we employ BM25 \cite{robertson2009bm25} to measure similarity between feedback texts. Since experience records preserve task-specific terminology and structural cues, lexical overlap provides an effective retrieval signal, making BM25 a lightweight yet effective choice that has also been shown to outperform embedding-based retrieval in software engineering scenarios \cite{yang2025empirical}.
Following \citet{zhao2025projectgen}, we set the maximum number of iterations to 3 for architecture design and skeleton generation stages, and to 5 for DevBench and 10 for CodeProjectEval in code implementation stage.

For both \ourmethod{} and baseline approaches, we ensure a fair comparison by providing the same set of inputs, including the PRD, UML diagrams, and architectural design documents. The PRD is a structured requirements document describing the system’s functional and non-functional aspects. UML diagrams (including class, package, and sequence diagrams) capture the structural relationships within the repository in Mermaid format. The architecture design document provides a high-level representation of the repository structure, including a source-code directory tree and concise descriptions of each file’s responsibilities, classes, and functions.

To ensure consistency and reproducibility, all experiments are conducted with \texttt{temperature=0}, while the agent responsible for generating code fix suggestions operates with \texttt{temperature=0.8}. The maximum context length is set to the upper limit of each LLM, with context windows of 128K tokens for DeepSeek-V3.2 and 400K tokens for GPT-5.

\section{Results}

\subsection{RQ1: Overall Performance}

We compare the performance of \ourmethod{} against state-of-the-art approaches on DevBench and CodeProjectEval. The results are presented in Table \ref{tab:main_results}.
We further analyze the results from three perspectives: functional correctness, architecture consistency, as well as code scale and complexity.

\subsubsection{Functional correctness}
As measured by the unit test pass rate (\%Pass), \ourmethod{} consistently achieves the best performance across all settings. On DevBench, it attains 50.81\% with DeepSeek-v3.2 and 60.48\% with GPT-5, outperforming all baselines by a clear margin. When evaluated on the more challenging CodeProjectEval dataset, the performance gap becomes even more pronounced. While CodeS and MetaGPT achieve competitive results on DevBench, their performance drops substantially on CodeProjectEval, indicating limited capability in handling complex project-level tasks. In contrast, \ourmethod{} remains the top performer, achieving 7.02\% and 27.73\%, demonstrating superior robustness and adaptability in more demanding scenarios.

The degradation of the baselines can be attributed to their design limitations. CodeS relies on a single-pass generation paradigm without iterative refinement or explicit quality control, making it susceptible to error propagation. Errors introduced in early stages cascade throughout the generation process, often resulting in file name resolution errors, incorrect code insertion, and unannotated natural language in generated files. MetaGPT directly generates code from natural language requirements without an explicit architectural design phase, leading to incomplete project structures, missing components, and partially defined function attributes. Although ProjectGen introduces an architecture design stage, its architecture is static and only serves as initialization for downstream generation. It also lacks mechanisms for accumulating abstract knowledge across iterations, limiting the refinement effectiveness.

\subsubsection{Architecture Consistency}
As CodeS and MetaGPT do not provide an explicit structural representation, we do not compute architecture consistency for them.
The architecture consistency results indicate that \ourmethod{} consistently produces repository structures that are more closely aligned with the ground-truth architecture than ProjectGen across different datasets and backbone models. 

ProjectGen relies solely on LLM-based judgment to evaluate architectural correctness, requiring the model to simultaneously reason over input specifications and generated architectural artifacts. Such long-context reasoning may reduce evaluation accuracy, particularly when complex structural dependencies need to be identified and compared. In contrast, \ourmethod{} augments LLM-based judgment with specialized tools that provide explicit structural evidence, enabling more accurate architecture verification and refinement.
This improvement is particularly important as architecture establishes the structural foundation for subsequent code generation. Higher architectural consistency provides stronger dependency priors and more reliable intermediate states, thereby facilitating more efficient iterative refinement and ultimately improving repository-level generation quality.

\subsubsection{Code scale and complexity}
From this perspective, larger code size or higher complexity does not necessarily imply better performance. A more desirable property is the ability to adapt the scale of generated code to the difficulty of the task. In this regard, \ourmethod{} produces solutions on DevBench with moderate lines of code and complexity, achieving strong performance while avoiding unnecessary overhead. On the more challenging CodeProjectEval dataset, \ourmethod{} generates substantially larger and more complex codebases, aligning with the increased structural demands of project-level tasks. Moreover, the scale of its generated codebases closely aligns with that of the ground-truth implementations across datasets. This behavior suggests that \ourmethod{} can effectively adapt the granularity and richness of generated programs to task difficulty.
By comparison, CodeS exhibits clear limitations in handling complex scenarios. On CodeProjectEval, it produces significantly smaller codebases in terms of both lines of code and number of files, which likely leads to insufficient functional coverage and contributes to its lower pass rates. This indicates that CodeS tends to under-generate for complex tasks and lacks the ability to appropriately scale its outputs. Notably, under the DeepSeek-v3.2 backbone on CodeProjectEval, the median performance of CodeS is 0, indicating that fewer than half of the generated projects can even be successfully parsed. On the DevBench dataset, where CodeS generates the largest and most complex codebases, it still achieves the lowest pass rates, suggesting that merely increasing code scale and complexity does not translate into better functional correctness.

\vspace{2mm}
\begin{custommdframed}
\textbf{Answer to RQ1:} \ourmethod{} achieves state-of-the-art performance across small- and large-scale project-level code generation tasks, demonstrating robust effectiveness and generalization across diverse backbones and datasets.
\end{custommdframed}

\subsection{RQ2: Ablation Study}

\begin{table}[t]
    \centering
    \setlength{\abovecaptionskip}{0.1cm}
    \caption{Abaltion results of \ourmethod{} on CodeProjectEval under GPT-5. \%Pass represents the overall pass rate on the dataset, while the other metrics are averaged across individual tasks.}
    \resizebox{\linewidth}{!}{
    \begin{tabular}{lccccc}
        \toprule
          & \textbf{\#Files} & \textbf{LOC} & \textbf{Complexity} & \textbf{Arch Cons.} & \textbf{\%Pass} \\
        \midrule
        w/o tools & 12.3 & 2371.7 & 5.80 & 81.44 & 10.88\% \\
        w/o co-evolution & 12.5 & 2298.7 & 5.35 & 82.28 & 21.08\%\\
        w/o experience & 13.3 & 2470.1 & 5.91 & 85.90 & 17.23\% \\
        \textbf{\ourmethod{}} & 13.3 & 2352.4 & 5.70 & \textbf{86.04} & \textbf{27.73\%}\\
        \bottomrule
    \end{tabular}
    }
    \label{tab:ablation}
    \vspace{-3mm}
\end{table}

\begin{figure}[t]
    \centering
    \setlength{\abovecaptionskip}{0.1cm}
    \includegraphics[width=\linewidth]{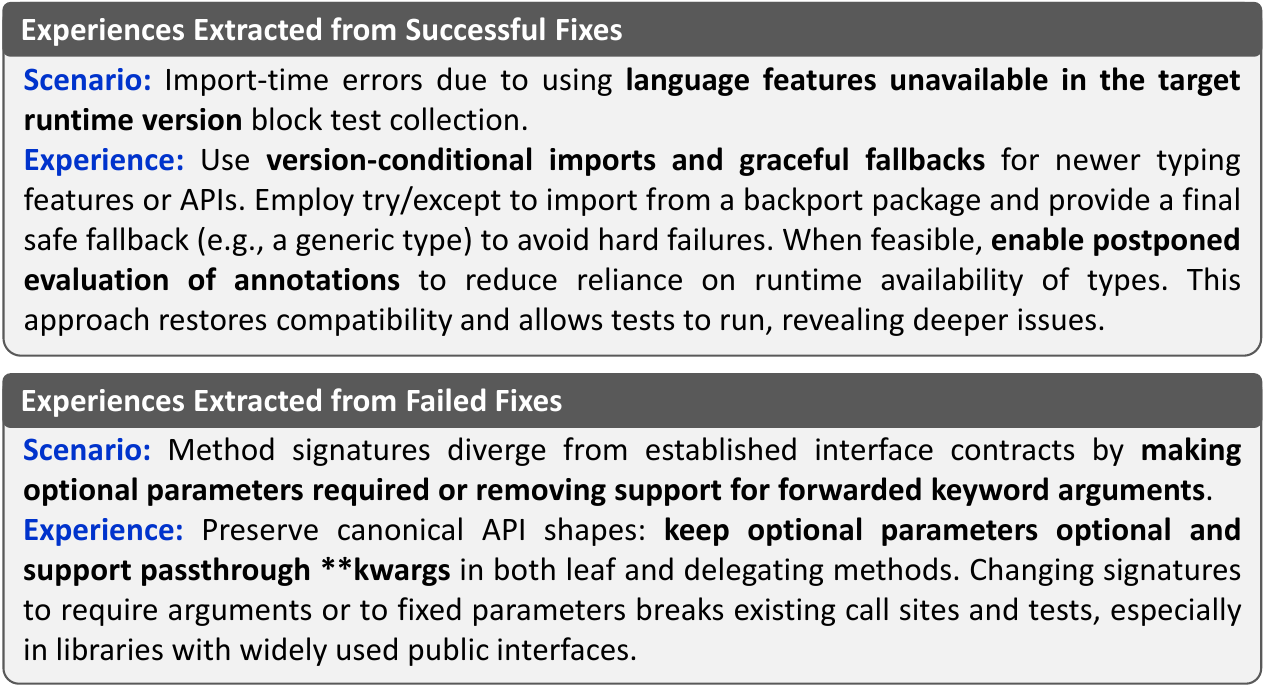}
    \caption{Examples of experience accumulated during iterative code generation.}
    \label{fig:experience}
    \vspace{-0.5cm}
\end{figure}

We conducted ablation experiments to evaluate the contributions of key components in \ourmethod{}. For the tools and experience modules, we completely removed the corresponding functionalities. For the architecture-code co-evolution mechanism, we limited the context construction to only the structural nodes generated during the initial architecture stage and the predicted relations produced in the skeleton stage, omitting the iterative co-evolution.
As shown in Table \ref{tab:ablation}, removing any single component leads to a clear decline in functional correctness as measured by pass rate, highlighting the complementary roles of these components.

Specifically, removing task-oriented tools results in a substantial decrease in pass rate from 27.73\% to 10.88\%, accompanied by the largest drop in architecture consistency. This indicates that tool support is critical not only for providing structured signals and reducing error propagation during generation, but also for enabling more accurate architecture verification and precise retrieval of relevant information necessary for code construction.

Similarly, eliminating the architecture-code co-evolution mechanism reduces the pass rate to 21.08\%, while causing only a moderate drop in architecture consistency. This suggests that the primary benefit of co-evolution lies not in improving the initial architecture itself, but in dynamically identifying relevant contextual dependencies and filtering out unrelated information during code generation, which is crucial for producing better outputs. Based on our calculations, the average project size on CodeProjectEval under GPT-5 is 19,376.06 tokens, while the context retrieved via the co-evolution mechanism has an average length of 6,560.93 tokens, accounting for only 33.9\% of the total. This demonstrates that co-evolution effectively focuses on the most relevant portions of the project, providing concise yet informative context that is sufficient to guide accurate code generation without the overhead of processing the entire repository.

The removal of experience-based memory also leads to a clear degradation, with the pass rate dropping to 17.23\%, while architecture consistency remains largely unchanged. This suggests that it mainly contributes to iterative refinement in detail rather than broad architecture construction. As illustrated in Figure \ref{fig:experience}, the retained experience consists of abstract and reusable insights that provide guidance during iterative refinement and help avoid repeating prior mistakes.

Additionally, the consistent code scale and complexity across all variants indicate that the overall architecture of \ourmethod{} is well-suited for generating substantial project code, while the individual components primarily contribute to improving functional correctness rather than altering code size or structural richness.

\vspace{2mm}
\begin{custommdframed}
\textbf{Answer to RQ2:} Each component of \ourmethod{} contributes to performance, with task-oriented tools providing the largest gains, while co-evolution and experience-based memory further improve robustness and reliability.
\end{custommdframed}

\subsection{RQ3: Efficiency Analysis}

\begin{figure}[t]
    \centering
    \setlength{\abovecaptionskip}{0.1cm}
    \begin{subfigure}[b]{0.48\linewidth}  
        \centering
        \includegraphics[width=\textwidth]{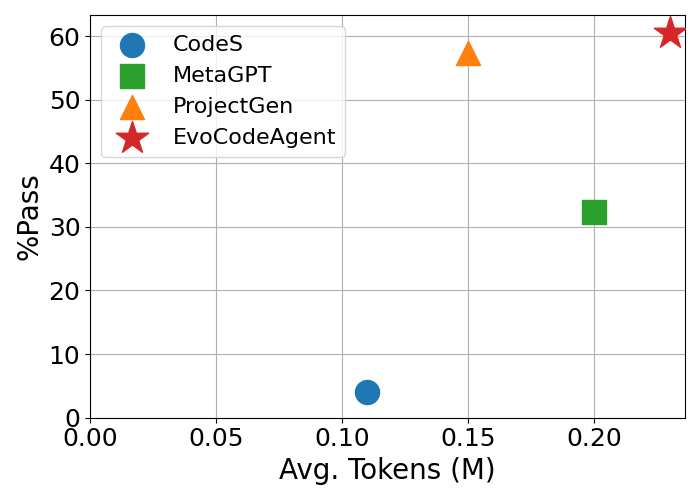} 
        \caption{DevBench}
        \label{fig:subfigA}
    \end{subfigure}
    \hfill
    \begin{subfigure}[b]{0.48\linewidth}
        \centering
        \includegraphics[width=\textwidth]{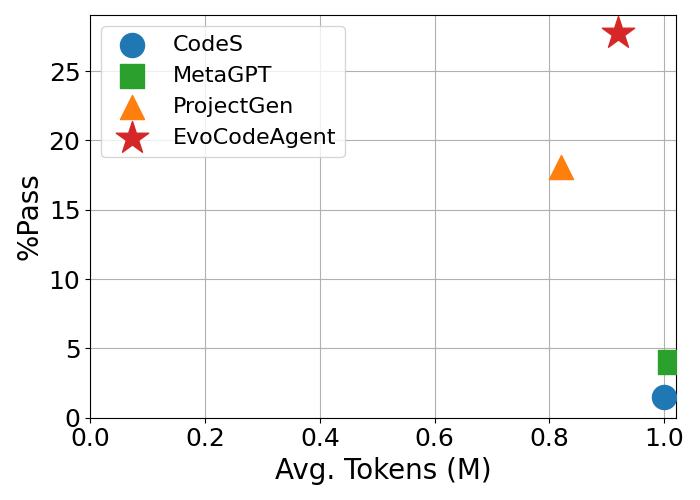}
        \caption{CodeProjectEval}
        \label{fig:subfigB}
    \end{subfigure}

        \caption{Comparison of \ourmethod{} and baselines on token consumption and performance under GPT-5.}
    \label{fig:cost}
    \vspace{-0.5cm}
\end{figure}

Figure \ref{fig:cost} presents an analysis of \ourmethod{} and baseline approaches, evaluating the trade-off between token consumption and generation performance under GPT-5. 

On the small-scale project-level code generation dataset DevBench, \ourmethod{} consumes approximately 1.2 times the tokens of MetaGPT, yet achieves nearly twice the unit test pass rate. Due to its use of tool calls and experience summarization, \ourmethod{} incurs additional LLM invocations and a slightly higher token cost compared to ProjectGen. Nevertheless, it converts this computational overhead into higher functional code quality.

This advantage of \ourmethod{} becomes more pronounced on the larger-scale project-level code generation dataset CodeProjectEval, which imposes stricter demands on long-context modeling and systematic code generation. While CodeS and MetaGPT consume the most tokens, their pass rates remain quite low. In contrast, \ourmethod{} achieves approximately 1.5 times the functional performance of ProjectGen with only a modest increase in token expenditure, highlighting its superior ability to scale efficiently to more complex projects. Notably, CodeS generates code on a per-function basis, providing all context at each step, which is impractical for large-scale projects, as it leads to excessively long contexts from which the LLM struggles to extract useful information. 

Across both datasets, \ourmethod{} consistently delivers strong generation performance at a competitive token cost. On CodeProjectEval, end-to-end project generation takes 70.6 minutes on average and incurs an estimated API cost of \$3.17 per repository. These results demonstrate that \ourmethod{}’s evolutionary agent-based framework effectively produces reliable and functionally correct code across a wide range of project-level scenarios, with its advantages becoming more pronounced in challenging large-scale tasks.

\vspace{2mm}
\begin{custommdframed}
\textbf{Answer to RQ3:}  \ourmethod{} achieves strong generation performance at a competitive token cost across both small- and large-scale datasets, with efficiency gains more pronounced on large-scale projects.
\end{custommdframed}

\begin{table}[t]
\centering
\caption{Statistics of experience on CodeProjectEval.}
\label{tab:experience_stats}
\small
    \begin{tabular}{llc}
    \toprule
    \textbf{Category} & \textbf{Metric} & \textbf{Value} \\
    \midrule
    \multirow{2}{*}{\textbf{Usage}} & Average experiences per task & 26.4 \\
     & Retrieval rate & 31.9\% \\
    \midrule
    \multirow{3}{*}{\textbf{Creation Stage}} & Early-stage experiences & 47.4\% \\
     & Middle-stage experiences & 31.1\% \\
     & Late-stage experiences & 21.5\% \\
    \midrule
    \multirow{3}{*}{\textbf{Retrieval Outcome}} & Improvement after retrieval & 35.3\% \\
     & No observable effect & 61.8\% \\
     & Regression after retrieval & 2.9\% \\
    \bottomrule
    \end{tabular}
    \vspace{-0.5cm}
\end{table}

\section{Discussion}

\subsection{Statistics and Effectiveness of Experience}

To better understand the role of abstracted experience, we analyze its usage statistics on CodeProjectEval, as summarized in Table \ref{tab:experience_stats}. A substantial portion of accumulated experiences are successfully retrieved, with nearly half of reusable experiences originating from early development stages. This suggests that useful development patterns are often discovered early and remain valuable throughout later iterations.

Retrieved experiences generally provide effective and safe auxiliary guidance. A considerable fraction of retrieval events leads to measurable improvements, while regression remains rare, indicating that experience reuse is unlikely to mislead the generation process. The effectiveness of retrieval largely depends on the similarity between current failures and previously encountered cases.

Qualitatively, retrieved experiences often capture recurring software engineering patterns, such as data validation, API consistency, serialization, and exception handling. These issues frequently involve cross-file dependencies and repeated implementation logic, making them particularly suitable for reuse across iterative refinement.

\subsection{Tool Usage Analysis}

\begin{figure}[t]
    \centering
    \includegraphics[width=\linewidth]{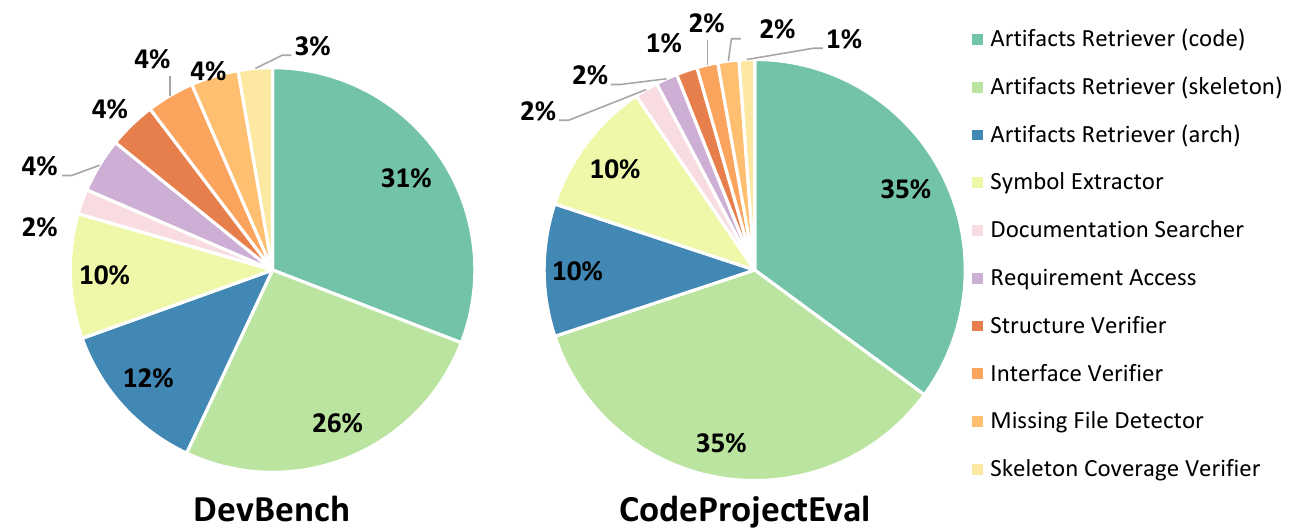}
    \caption{Distribution of tool invocation frequencies of \ourmethod{} under GPT-5.}
    \label{fig:tool}
    \vspace{-0.5cm}
\end{figure}

To investigate tool usage across projects of different scales, we counted tool invocations on DevBench and CodeProjectEval, as shown in Figure \ref{fig:tool}.

Across both datasets, artifact retrieval tools dominate overall usage, with code and skeleton retrieval accounting for the largest proportion, highlighting the importance of structured access to intermediate artifacts for maintaining cross-file consistency. Symbol extraction and architecture retrieval also remain stable, indicating that structure-aware reasoning is essential regardless of project scale.
As task scale increases, tool usage becomes more concentrated. On the smaller-scale DevBench dataset, tool usage is more diverse, with verification and auxiliary tools contributing noticeably. In contrast, on CodeProjectEval, code and skeleton retrieval occupy a much larger proportion, suggesting that large-scale generation depends increasingly on efficient context retrieval, while the relative role of verification diminishes.

\subsection{Case Study}

To better understand the behavioral differences between workflow-based software engineering agents and monolithic CLI-based coding agents, we conduct a case study using Claude Code\footnote{\url{https://www.claude.com/product/claude-code}} as a representative CLI-based baseline on the \texttt{deprecated} project from CodeProjectEval. Claude Code receives the same structured inputs as \ourmethod{}, including the PRD, UML diagrams, and architectural design documents. We evaluate it under two settings: \textit{self-testing}, where it autonomously generates and validates the entire project, and \textit{external-feedback}, where it generates only source code and receives check-test feedback after each iteration.

Table~\ref{tab:case} summarizes the results. Claude Code demonstrates strong end-to-end generation capability, benefiting from a unified conversational workflow that enables rapid development and flexible refinement. However, this design also leads to steadily growing context and higher token consumption during iterative revisions. In contrast, \ourmethod{} decomposes development into specialized stages with independent contexts, enabling more efficient feedback-driven refinement. These results highlight a trade-off between the flexibility of unified long-context generation and the efficiency of modular workflow-based software engineering.

\begin{figure}
    \centering
    \includegraphics[width=\linewidth]{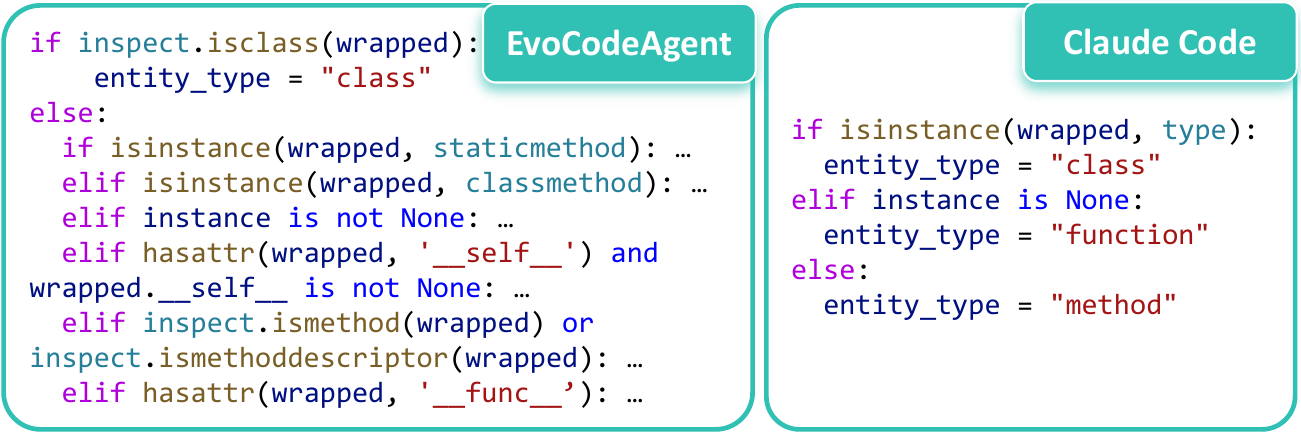}
    \caption{Comparison of callable type identification for deprecation message generation.}
    \label{fig:case}
    \vspace{-0.5cm}
\end{figure}

Figure~\ref{fig:case} illustrates a representative implementation difference. Claude Code adopts a simpler invocation-based strategy for callable classification, which offers lower implementation complexity and is often sufficiently robust for common cases. In contrast, \ourmethod{} performs finer-grained analysis through Python introspection. In this example, execution feedback enables \ourmethod{} to identify and correct the mismatch during subsequent iterations, illustrating how structured decomposition improves error localization and recovery. This comparison indicates the trade-off between lightweight heuristic-based generation and more structured but analysis-intensive workflow design.

\begin{table}[t]
    \centering
    \setlength{\abovecaptionskip}{0.1cm}
    \caption{Comparison between \ourmethod{} and Claude Code on \texttt{deprecated} project under DeepSeek-v3.2.}
    \resizebox{\linewidth}{!}{
    \begin{tabular}{lcccc}
        \toprule
        \textbf{Approach}& \textbf{\%Pass} & \textbf{Tokens} \\
        \midrule
        \textbf{ClaudeCode$_{\text{self-testing}}$} & 11.2\% & 2,244,863\\
        \textbf{ClaudeCode$_{\text{external-feedback}}$} & 18.8\% & 2,203,084\\
        \textbf{\ourmethod{}} & 51.4\% & 685,589\\
        \bottomrule
    \end{tabular}
    }
    \label{tab:case}
    \vspace{-0.3cm}
\end{table}

\subsection{Threats to Validity}

The primary threats to validity stem from dataset coverage, backbone model selection, and input quality. First, our experiments are conducted on a limited set of datasets, which may constrain the generalizability of the results. To mitigate this threat, we evaluate \ourmethod{} on two benchmarks of different scales, covering both small-scale and large-scale project-level code generation, where \ourmethod{} consistently outperforms baseline methods. Second, model-specific bias may affect performance. To reduce this risk, we evaluate \ourmethod{} with two representative LLMs and observe consistent performance gains across both settings. A further threat arises from input quality, as incomplete or contradictory requirements may degrade generation quality. To alleviate this issue, \ourmethod{} employs closed-loop feedback with iterative project state refinement and leverages multiple input artifacts to compensate for missing information. Finally, \ourmethod{} is inherently language-agnostic, and its core mechanisms can be extended to other programming languages in future work.

\section{Conclusion}

In conclusion, this work presents \ourmethod{}, a multi-agent framework for project-level code generation that effectively addresses key limitations of existing approaches. 
By combining an evolving project state, tool-grounded verification, and iterative experience accumulation, \ourmethod{} enables reliable state reconciliation and continuous self-improvement.
Experimental results on diverse benchmarks, including DevBench and CodeProjectEval, show that \ourmethod{} consistently surpasses state-of-the-art methods, achieving substantial improvements in unit test pass rates and demonstrating robust generalization across projects of varying complexity. These findings highlight the potential of combining iterative, experience-aware generation with multi-agent collaboration and dynamic architecture-code co-evolution to enhance the reliability, consistency, and scalability of automated software synthesis.

\bibliographystyle{IEEEtranN}
\footnotesize
\bibliography{references}

\end{document}